# Instrumental distortions in quantum optimal control

Uluk Rasulov[1] and Ilya Kuprov[2,1,*]

[1]*School of Chemistry and Chemical Engineering, University of Southampton, United Kingdom.*

[2]*Department of Chemical and Biological Physics, Weizmann Institute of Science, Israel.*

## Abstract

Quantum optimal control methods, such as gradient ascent pulse engineering (GRAPE), are used for precise manipulation of quantum states. Many of those methods were pioneered in magnetic resonance spectroscopy where instrumental distortions are often negligible. However, that is not the case elsewhere: the usual jumble of cables, resonators, modulators, splitters, amplifiers, and filters can and would distort control signals. Those distortions may be non-linear, their inverse functions may be ill-defined and unstable; they may even vary from one day to the next, and across the sample.

Here we introduce the response-aware gradient ascent pulse engineering (RAW-GRAPE) framework, which accounts for any cascade of differentiable distortions within the GRAPE optimisation loop, does not require filter function inversion, and produces control sequences that are resilient to user-specified distortion cascades with user-specified parameter ensembles.

The framework is implemented into the optimal control module supplied with versions 2.10 and later of the open-source *Spinach* library; the user needs to provide function handles returning the actions by the distortions and, optionally, parameter ensembles for those actions.

*Email: ilya.kuprov@weizmann.ac.il

## 1. Introduction

Gradient ascent pulse engineering (GRAPE)[1] and its extensions[2-5] are currently the dominant quantum optimal control frameworks in magnetic resonance spectroscopy[6,7] and imaging[8], and increasingly in other quantum process control applications[9], a notable recent example being atom interferometry[10,11]. All published GRAPE implementations assume that control sequence distortions by instrument filter functions are negligible. With one recent exception[5], all implementations also use the piecewise-constant approximation for the control Hamiltonian without worrying about edge transients. In high-field liquid state nuclear magnetic resonance (NMR) those are indeed negligible[12], but elsewhere in quantum device engineering that is not the case[13].

The reason why minor distortions can be ignored is the definition of the GRAPE optimum: a zero gradient of the fidelity with respect to the control sequence[1] implies first order resilience to variations thereof. In benign cases (magnetic resonance[6,14], atom interferometry[10,11]), optimal control theory yields impressive results right out of the box. However, in low-γ NMR and in time domain electron spin resonance (ESR) hardware distortions of the control sequence can be significant[13,15-20] and must be taken into account. The difficulty is that instrumental filter functions are hard to measure[13,16,20-24] and do not usually have a well-defined inverse. The problem of creating a control sequence that is distorted into the desired sequence is therefore ill-posed[18-20,25-27].

Here we point out that instrument filter functions do not need to be inverted or even measured precisely – we implement an extension of the GRAPE formalism that incorporates ensembles of instrument filter function cascades with user-specified parameter ranges into the optimisation loop. In this approach, only the forward action by the distortion and the Jacobian of that action are needed; both are well defined and stable. Control sequences may now be designed to be stable to multiple types of individual and sequentially applied distortions and distributions in their parameters. We call this formalism response-aware gradient ascent pulse engineering (RAW-GRAPE) and report its implementation in versions 2.10 and later of the open-source *Spinach* library[28].

## 2. Dissipative GRAPE framework

Decoherence and return to thermal equilibrium are unavoidable; this is the principal failure modality for quantum devices. Any optimal control framework must therefore take dissipative processes into account by construction; here we use the Liouville-space (*aka* adjoint representation) version of the GRAPE framework[29,30], the unitary case is recovered by setting the relaxation superoperator to zero. In magnetic resonance notation, the general equation of motion is[31]:

$$\frac{d}{dt}\boldsymbol{\rho}(t) = -i\mathcal{L}(t)\boldsymbol{\rho}(t), \quad \mathcal{H}\boldsymbol{\rho} = [\mathbf{H},\boldsymbol{\rho}]$$
$$\mathcal{L}(t) = \mathcal{H}(t) + i\mathcal{F}(t) + i\mathcal{K}(t) + i\mathcal{R}(t)$$

(1)

where $\boldsymbol{\rho}$ is the density matrix, $\mathcal{H}(t)$ is the spin Hamiltonian commutation superoperator, $\mathcal{F}(t)$ is the diffusion and flow generator, $\mathcal{K}(t)$ is the chemical kinetics superoperator, and $\mathcal{R}(t)$ is the relaxation superoperator in which time dependence is uncommon, but may be present, for example, when the main magnet field is a function of time in low-field NMR spectroscopy and relaxometry[32].

GRAPE algorithm splits the Liouvillian $\mathcal{L}(t)$ into the uncontrollable "drift" generator $\mathcal{D}(t)$ and a linear combination of control generators $\mathcal{C}_k$ whose coefficients $c^{(k)}(t)$ the instrument can vary:

$$\mathcal{L}(t) = \mathcal{D}(t) + \sum_k c^{(k)}(t)\mathcal{C}_k \qquad (2)$$

In the context of magnetic resonance spectroscopy and imaging, $\mathcal{D}(t)$ may contain Zeeman interactions with the main magnet field, spin-spin couplings, diffusion and hydrodynamics[33], magic angle spinning[34], decoherence[35], and other processes that are beyond the possibility of agile variation. The control part may contain interactions with the fields generated by radiofrequency coil arrays, microwave resonators, and pulsed field gradient coil arrays.

Consider an experiment of duration $T$ with the user-specified set of initial conditions $\{\boldsymbol{\rho}_0^{(m)}\}$ that must be brought into the corresponding set of desired destination states $\{\boldsymbol{\delta}^{(m)}\}$. Depending on the size of this set, the problem may be of state-to-state type (single source and destination), subspace-to-subspace type (multiple source-destination pairs), and gate design type (complete linearly independent set of source-destination pairs). In all three cases, popular fidelity measures are functions of the overlap $\Omega$ between each evolved initial condition and the corresponding target state:

$$\Omega\{\boldsymbol{\rho},\boldsymbol{\delta},\mathbf{C}\} = \langle\boldsymbol{\delta}|\overleftarrow{\exp}\left[-i\int_0^T \mathcal{L}(t)dt\right]|\boldsymbol{\rho}\rangle$$

$$\overleftarrow{\exp}\left(-i\int_0^t \mathcal{L}(t)dt\right) = \lim_{\Delta t_n \to 0} \overleftarrow{\prod_n} \exp\left[-i\mathcal{L}(t_n)\Delta t_n\right] \qquad (3)$$

where the arrow indicates Dyson's time-ordered exponential[36], defined as the limit of a time-ordered product, and $\mathbf{C}$ is the set of control sequences defined in Eq (2). GRAPE discretises this problem in time[1], the simplest case uses a set $\{\Delta t_n\}$ of sufficiently short time intervals to approximate the drift generator and the control sequences as piecewise-constant:

$$\Omega\{\boldsymbol{\rho},\boldsymbol{\delta},\mathbf{C}\} = \langle\boldsymbol{\delta}|\mathcal{P}_N \cdots \mathcal{P}_{n+1}\mathcal{P}_n\mathcal{P}_{n-1}\cdots\mathcal{P}_1|\boldsymbol{\rho}\rangle$$

$$\mathcal{P}_n = \exp\left[-i\left(\mathcal{D}_n + \sum_k c_n^{(k)}\mathcal{C}_k\right)\Delta t_n\right] \qquad (4)$$

$$c^{(k)}(t) = c_n^{(k)}, \qquad \mathcal{D}(t) = \mathcal{D}_n, \qquad t_{n-1} \le t < t_n$$

For historical reasons, optimisation modules of technical computing software are programmed to perform minimisation. Software implementations of optimal control algorithms therefore commonly define "infidelity" as the difference between some theoretical best value of the figure of merit and its current value. What enables efficient optimisation (by steepest descent[37], conjugate gradients[38], or quasi-Newton methods[39]) is the gradient of $\Omega$ with respect to the control sequence. GRAPE is popular because that gradient is surprisingly cheap[2] – for each control channel $k$ at each time step $\Delta t_n$ only one action by the propagator derivative on a state vector is needed:

$$\frac{\partial \Omega}{\partial c_n^{(k)}} = \langle\boldsymbol{\delta}|\mathcal{P}_N \cdots \mathcal{P}_{n+1} \frac{\partial \mathcal{P}_n}{\partial c_n^{(k)}} \mathcal{P}_{n-1} \cdots \mathcal{P}_1|\boldsymbol{\rho}\rangle \qquad (5)$$

The equation for the matrix exponential derivative is memorably elegant[40]:

$$\exp\left[\begin{pmatrix} \mathbf{A} & \partial\mathbf{A}/\partial\alpha \\ \mathbf{0} & \mathbf{A} \end{pmatrix}\right] = \begin{pmatrix} e^{\mathbf{A}} & \partial e^{\mathbf{A}}/\partial\alpha \\ \mathbf{0} & e^{\mathbf{A}} \end{pmatrix} \qquad (6)$$

As a result, the action by $\partial e^{\mathbf{A}}/\partial\alpha$ on a vector $\mathbf{x}$ may be computed simultaneously with the action by $e^{\mathbf{A}}$ using the following block matrix relation[41]:

$$\begin{pmatrix} \left[\partial e^{\mathbf{A}}/\partial\alpha\right]\mathbf{x} \\ e^{\mathbf{A}}\mathbf{x} \end{pmatrix} = \begin{pmatrix} e^{\mathbf{A}} & \partial e^{\mathbf{A}}/\partial\alpha \\ \mathbf{0} & e^{\mathbf{A}} \end{pmatrix}\begin{pmatrix} \mathbf{0} \\ \mathbf{x} \end{pmatrix} = \\ = \exp\left[\begin{pmatrix} \mathbf{A} & \partial\mathbf{A}/\partial\alpha \\ \mathbf{0} & \mathbf{A} \end{pmatrix}\right]\begin{pmatrix} \mathbf{0} \\ \mathbf{x} \end{pmatrix} \qquad (7)$$

Such products may be efficiently calculated without explicitly exponentiating the matrix[42,43]. The derivative of the evolution generator $\mathbf{A}$ is simply a multiple of the control operator:

$$\frac{\partial}{\partial c_n^{(k)}}\left[-i\left(\mathcal{D}_n + \sum_k c_n^{(k)}\mathcal{C}_k\right)\Delta t_n\right] = -i\mathcal{C}_k\Delta t_n \qquad (8)$$

Dozens of case-specific corners may be cut at various levels of this procedure to improve the computational efficiency, sometimes by orders of magnitude[4,44-51]. At the time of writing, the most general and flexible implementation is *Spinach*[28]; the fastest no-frills implementation (single-spin Bloch equations with phase-modulated controls and without relaxation) is *SEEDLESS*[46]. A vibrant industry exists; GRAPE is a common straw to clutch at when a quantum computing architecture fails.

## 3. Distortion cascade integration into GRAPE

Consider a cascade of instrumental distortions $\rightarrow \mathbf{Q}^{(1)} \rightarrow \mathbf{Q}^{(2)} \rightarrow ... \rightarrow \mathbf{Q}^{(J)} \rightarrow$, caused by every physically distinct item of hardware in the control circuit and represented here by continuous and differentiable filter functions $\mathbf{Q}^{(j)}(\mathbf{C})$. Each of them takes a control coefficient array as input, and returns another as an output; the layout used by *Spinach* has channels (for example X and Y magnetic fields) along the vertical dimension and time slices along the horizontal dimension. Thus, each filter function $\mathbf{Q}^{(j)}(\mathbf{C})$ takes a matrix and returns another matrix. All control channels must be supplied simultaneously because channel cross-talk is a common type of instrumental distortion.

Our task is to introduce this distortion cascade into the GRAPE optimisation loop for an ensemble of distortion parameter values; the ensemble must reflect practical variations in those parameters between instruments, samples, and measurement sessions. We assume that $\{\mathbf{Q}^{(j)}\}$ are easily computable and differentiable but their inverses may be ill-posed and unstable.

The procedure is analytically straightforward – the figure of merit is modified to have the control sequences affected by each distortion function before acting on the system:

$$\Omega(\mathbf{C}) \quad \rightarrow \quad \Omega\left(\mathbf{Q}^{(J)}\left(...\mathbf{Q}^{(2)}\left(\mathbf{Q}^{(1)}(\mathbf{C})\right)\right)\right) \qquad (9)$$

These are simply nested functions – derivatives of $\Omega$ with respect to control coefficients must therefore be modified by the multivariate chain rule:

$$\frac{\partial \Omega}{\partial c_n^{(k)}} \quad \rightarrow \quad \sum_{pqij\ldots lmrs} \frac{\partial \Omega}{\partial q_q^{(J,p)}} \frac{\partial q_q^{(J,p)}}{\partial q_j^{(J-1,i)}} \cdots \frac{\partial q_m^{(2,l)}}{\partial q_s^{(1,r)}} \frac{\partial q_s^{(1,r)}}{\partial c_n^{(k)}} \tag{10}$$

where $q_c^{(a,b)}$ is the output of the distortion function $\mathbf{Q}^{(a)}$ on control channel $b$ at time point $c$, and therefore the derivative $\partial q_c^{(a,b)} / \partial q_{c'}^{(a-1,b')}$ is a matrix element of the Jacobian – the slope of the response of each output element $q_c^{(a,b)}$ of $\mathbf{Q}^{(a)}$ with respect to the change in each output element $\partial q_{c'}^{(a-1,b')}$ of the preceding distortion $\mathbf{Q}^{(a-1)}$. The modified fidelity gradient in Eq (10) has a more eye-friendly schematic expression in the matrix calculus notation:

$$\frac{\partial \Omega}{\partial \mathbf{C}} \quad \rightarrow \quad \frac{\partial \Omega}{\partial \mathbf{Q}^{(J)}} \cdot \frac{\partial \mathbf{Q}^{(J)}}{\partial \mathbf{Q}^{(J-1)}} \cdots \frac{\partial \mathbf{Q}^{(2)}}{\partial \mathbf{Q}^{(1)}} \cdot \frac{\partial \mathbf{Q}^{(1)}}{\partial \mathbf{C}} \tag{11}$$

Each dot product here contracts a pair of indices; our *Matlab* implementation is located in `ensemble.m` function of *Spinach*, its unit tests are in the example set (`fundamentals/derivative_tests`).

The chain of Jacobians in Eqs (10) and (11) is identical to the one found in backpropagation training of deep neural networks[52] – in fact, that algorithm may be re-used line for line: the forward propagation step is Eq (9), then the gradient with respect to the output of the last filter function $\partial \Omega / \partial \mathbf{Q}_J$, and then the chain of Jacobian multiplications in Eq (11) is the backpropagation. The difference with neural networks is that we do not optimise distortion parameters.

The user needs to supply function handles for the filter functions $\{Q_j\}$ and their Jacobians. Alternatively (at the cost of computational efficiency), Jacobians may be obtained using automatic differentiation. Note that this approach does not require inverting any of the (in general, non-invertible) filter functions. Thus, a complication present in some of the earlier methods dealing with instrumental distortion incorporation into GRAPE[16,18-20] is eliminated.

### 3.1 Jacobians of linear distortions

Consider first the well-researched case[16] of a distortion representable by a linear filter with a memory kernel $h(t)$ where the input signal $u(t)$ starts at $t = 0$. The output $v(t)$ is:

$$v(t) = \int_0^t h(\tau) u(t-\tau) d\tau \tag{12}$$

When the uniform time discretisation and the piecewise-constant approximation matching Eqs (3) and (4) of the GRAPE algorithm are applied, the integral becomes (zero base indexing):

$$v_n = \sum_{m=0}^{n} h_m u_{n-m} \Delta t, \qquad u_{k<0} = 0 \tag{13}$$

where the vector $\mathbf{h}$ has a physical meaning of attenuation coefficients with which the system remembers its past. In terms of the input vector $\mathbf{u}$ and the output vector $\mathbf{v}$, a convenient formulation of Eq (13) is $\mathbf{v} = \mathbf{H}\mathbf{u}$ where $\mathbf{H}$ is a Toeplitz matrix constructed from $\mathbf{h}$:

$$\mathbf{H} = \begin{bmatrix} h_0 & 0 & 0 & \cdots & 0 \\ h_1 & h_0 & 0 & \cdots & 0 \\ h_2 & h_1 & h_0 & \cdots & 0 \\ \vdots & \vdots & \vdots & \cdots & \vdots \end{bmatrix} \Delta t \qquad (14)$$

Computing the derivatives $\partial v_n / \partial u_k$ then demonstrates that no more work is needed: this matrix is already the Jacobian of the distortion transformation[21,22,24]. The Toeplitz form in Eq (14) is also logistically convenient because the number of non-zero elements in $\mathbf{h}$ is usually small compared to the length of the input vector $\mathbf{u}$, and the matrix $\mathbf{H}$ is therefore sparse.

**3.2 Generator set for any linear distortion**

Many instrumental distortions may be described by linear filter functions. Conveniently, only two elementary filters are needed to generate an arbitrary linear distortion by cascades and linear combinations: a single-pole (SP) filter and a single-zero (SZ) filter, both with a unit DC gain[53]:

$$\begin{aligned} v_n^{\text{SP}} &= (1-p)u_n + pv_{n-1}^{\text{SP}} \\ v_n^{\text{SZ}} &= (u_n - zu_{n-1})/(1-z) \end{aligned} \qquad (15)$$

where $z$ and $p$ are dimensionless filter coefficients:

$$p = e^{-r_{\text{P}}\Delta t + i(\omega_{\text{P}} - \omega_{\text{RF}})\Delta t}, \qquad z = e^{-r_{\text{P}}\Delta t + i(\omega_{\text{Z}} - \omega_{\text{RF}})\Delta t} \qquad (16)$$

that depend on the damping rates $r_{\text{P,Z}}$ and frequencies $\omega_{\text{P,Z}}$ of the pole and the zero, and on the time discretisation step $\Delta t$. In rotating frames (for example, with heterodyne detection at a particular frequency $\omega_{\text{RF}}$), $\omega_{\text{P}}$ and $\omega_{\text{Z}}$ are shifted by that frequency.

With cascades of these filters implemented and their Jacobians chained as shown in Eq (11), this extension of GRAPE can therefore handle an arbitrary linear distortion. We have also implemented linear combinations of cascades and their ensembles with respect to distortion parameter variations.

Figure 1 shows examples of the action by a second-order low-pass filter made by cascading two single-pole filters, a third-order high-pass filter made by cascading three single-zero filters, and an underdamped RLC circuit distortion obtained by cascading two single-pole filters.

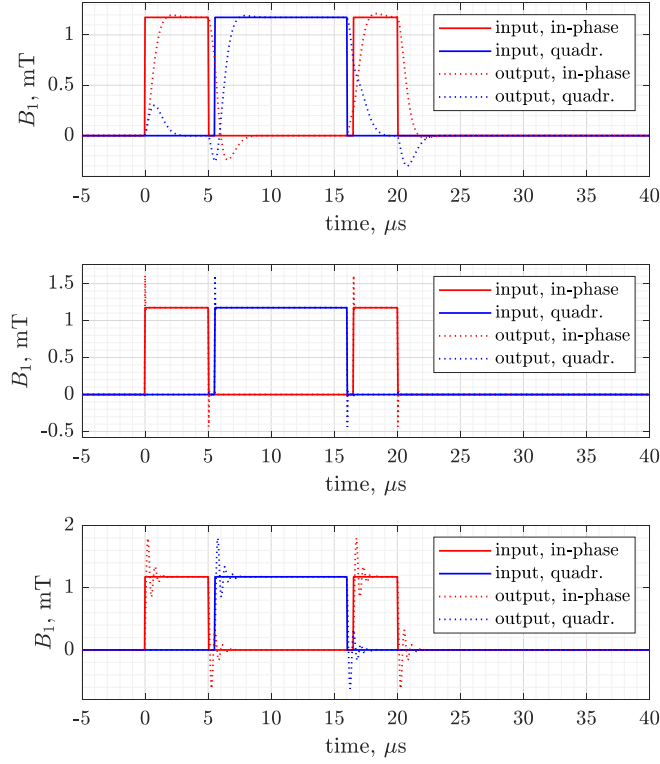

*Figure 1. Examples of the action by cascades of single-pole and single-zero filters on the popular $90_X180_Y90_X$ composite inversion pulse[54], with individual pulses slightly spaced out for visual clarity and $B_1$ field calibrated for proton Larmor frequency. **(Top Panel)** a cascade of two identical single-pole filters with the pole slightly off resonance relative to the transmitter frequency. **(Middle Panel)** a cascade of three identical single-zero filters with the zero slightly off resonance relative to the transmitter frequency. **(Bottom Panel)** an RLC circuit response filter implemented as a cascade of two single-pole filters with the opposite imaginary parts in the location of the two poles.*

### 3.3 RLC circuit distortion

A practically important linear filter that deserves a special mention is the distortion introduced by an RLC circuit[20,55] with a natural frequency $\omega$ and a quality factor $Q$:

$$\omega = \frac{1}{\sqrt{LC}}, \qquad Q = \frac{1}{R}\sqrt{\frac{L}{C}} \tag{17}$$

This distortion (illustrated in the bottom panel of Figure 1) may be factored into a cascade of two discrete single-pole filters with the following expression for the poles[56]:

$$p_{1,2} = \exp\left(-\frac{|\omega|}{2Q}\Delta t \pm i(\omega - \omega_{RF})\Delta t\sqrt{1-\frac{1}{4Q^2}}\right) \tag{18}$$

where $\Delta t$ is the time discretisation step and $\omega_{RF}$ is the rotating frame frequency. The damping rate in this case is determined by the ratio of the natural frequency and the quality factor.

### 3.4 More complex distortions

Filter functions of control hardware may be non-linear[23,24]. Amplifiers are one example; they saturate when operating near their maximum output power. This is a common enough scourge that magnetic resonance instrument manufacturers take special measures to make their amplifiers look linear to the user; a significant body of work exists on estimating filter functions of hardware components[13,16,22]. Two simple amplitude saturation models supplied with *Spinach* are:

$$v = a\tanh(u/a), \qquad v = x\Big/\sqrt[s]{1+(u/a)^s} \qquad (19)$$

where $u$ is the input amplitude, $v$ is the output amplitude, $a$ is the saturation level, and the sharpness of the transition from the linear to the saturating behaviour in the reciprocal root model is regulated by the parameter $s > 1$. Figure 2 shows the effect of amplifier saturation described by Eq (19) on a popular Veshtort-Griffin E1000B radiofrequency pulse[57] that performs band-selective excitation in magnetic resonance spectroscopy.

In general, non-linear distortions are hardware- and system-specific; a software implementation cannot anticipate any properties other than the model being differentiable. At the software level we have therefore opted for automatic differentiation: users need only to supply the function itself; the Jacobian is generated using the `dljacobian.m` function of *Matlab*'s Deep Learning Toolbox[58]. If computational efficiency is a concern, Jacobians may also be hand-coded.

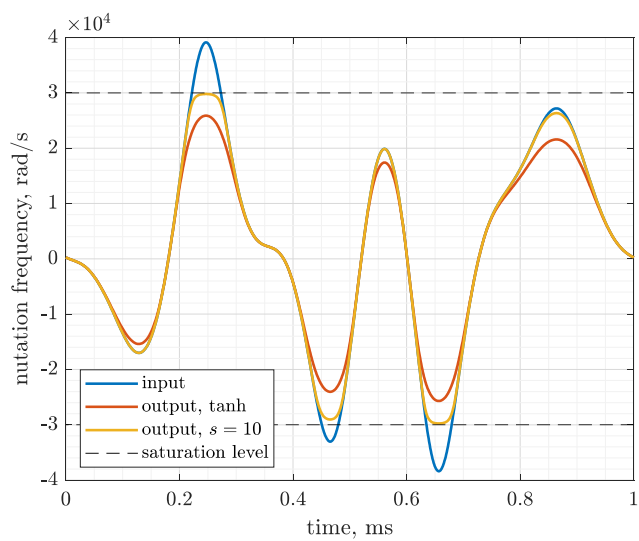

*Figure 2. An example of the amplifier compression effect, computed using Eq (19), on a Veshtort-Griffin E1000B radiofrequency pulse used in NMR spectroscopy to achieve band-selective magnetisation excitation[57].*

Common hardware non-linearities provided as templates with *Spinach* package[28] are:

**Phase distortions**. These originate from dispersive effects of filters, cables, and connectors, and from other hardware that introduces frequency-dependent phase shifts[59].

**Frequency-dependent attenuation**. Hardware components can attenuate specific frequencies; high frequencies are particularly vulnerable due to skin effects and dielectric losses[60].

**Nonlinear crosstalk.** This occurs when signals in spatially adjacent components interfere nonlinearly, producing intermodulation products; cross-talk is particularly problematic in compact systems and integrated circuits, where channels are hard to isolate[61].

**Signal timing errors**. Sample timing errors in digital signals can effectively shift frequencies and introduce irreproducible additional modulations[62].

**Digital quantisation**: Analogue-to-digital and digital-to-analogue converters approximate continuous signals using discrete levels; this quantisation can manifest as a noise source[56].

To generalise the procedures and functions above to arbitrary differentiable distortions, we use a well-known result from machine learning: the universal approximator theorem, which states that a sufficiently long superposition of linear transformations with element-wise non-linearities can approximate any Borel measurable map between vector spaces to any accuracy[63]. We therefore conclude that, at least in principle, any instrumental distortion can be accommodated within GRAPE using cascades, linear combinations, and ensembles of the components described above.

## 4. Performance illustrations

To illustrate the performance of the RAW-GRAPE framework, we consider universal rotation pulse design – a common problem in magnetic resonance[64] and atom interferometry[11]. We model typical hardware distortions seen in those spectroscopies: probe circuit ringing and amplifier saturation. As magnetic resonance fields get higher (at the time of writing, cutting-edge NMR instruments run at 28.18 Tesla), optimal control methods become unavoidable[14,44,64] because ideal pulses are no longer instrumentally achievable: commonly available RF nutation frequencies in latest cryoprobes at 28.18 Tesla are 20 kHz on $^1$H, 15 kHz on $^{13}$C, and 5 kHz on $^{15}$N. Throughout this section we use simulated performance profiles interchangeably with experimental data; this is warranted in magnetic resonance because simulation methods are exceptionally well developed and provide a close match.

Our universal rotation pulse must accomplish the following transformation of single-spin density matrices (for spin 1/2, these are Pauli matrices scaled to have $[\mathbf{S}_X, \mathbf{S}_Y] = i\mathbf{S}_Z$):

$$\mathbf{S}_Z \to \mathbf{S}_X, \quad \mathbf{S}_Y \to \mathbf{S}_Y, \quad \mathbf{S}_X \to -\mathbf{S}_Z \tag{20}$$

within a given chemical shift range (we chose 200 ppm for $^{13}$C in a 28.18 Tesla magnet). The pulse must be short enough for the worst-case $^{13}$C-$^1$H *J*-coupling (around 200 Hz) to have a negligible effect; this caps the duration at about 1/100*J* = 50 μs. Maximum instrumentally achievable $^{13}$C nutation frequency in a room temperature NMR probe varies from 50 to 70 kHz across the radiofrequency coil of the NMR probe, and therefore a hard $^{13}$C pulse (*i.e.* the shortest pulse at the maximum available RF power, here about 4 μs) yields significant phase errors across the spectral window (Figure 3, top left).

For room temperature NMR probes (Q-factors between 50 and 200), this can be overcome using optimal control (Figure 3, bottom left) with the additional advantage of lower pulse power[14,44,64]. However, stronger RLC circuit distortions render such pulses ineffectual: when the circuit quality factor is increased into thousands (corresponding to a narrowly optimised cryoprobe[55]), the fidelity of the pulse degrades (Figure 3, top right). At that point, introducing RLC distortion modelling into the GRAPE loop as described above eliminates the problem: when RAW-GRAPE is used, the performance of the RLC distorted pulse returns to the high fidelity (Figure 3, bottom right).

At the software implementation level, we have condensed the procedures described above into just one additional user input line in the optimal control module of the *Spinach* package[51]:

```
% Add RLC distortion to the GRAPE workflow
control.distortion = { @(w)spf(w,p1), @(w)spf(w,p2) };
```

Here, the distortion cascade is specified as a cell array of *Matlab* function handles. The functions are both SPF (single pole filter, supplied with *Spinach*) and their parameters $p_1$ and $p_1$ are the two filter coefficients from Eq (18), where $\omega$ is the Larmor frequency of $^{13}$C and the quality factor $Q$ is set by the user. A number of other distortion functions are provided with *Spinach* kernel, and more can be added by the user in a matter of minutes because *Matlab* takes care of the Jacobian calculation.

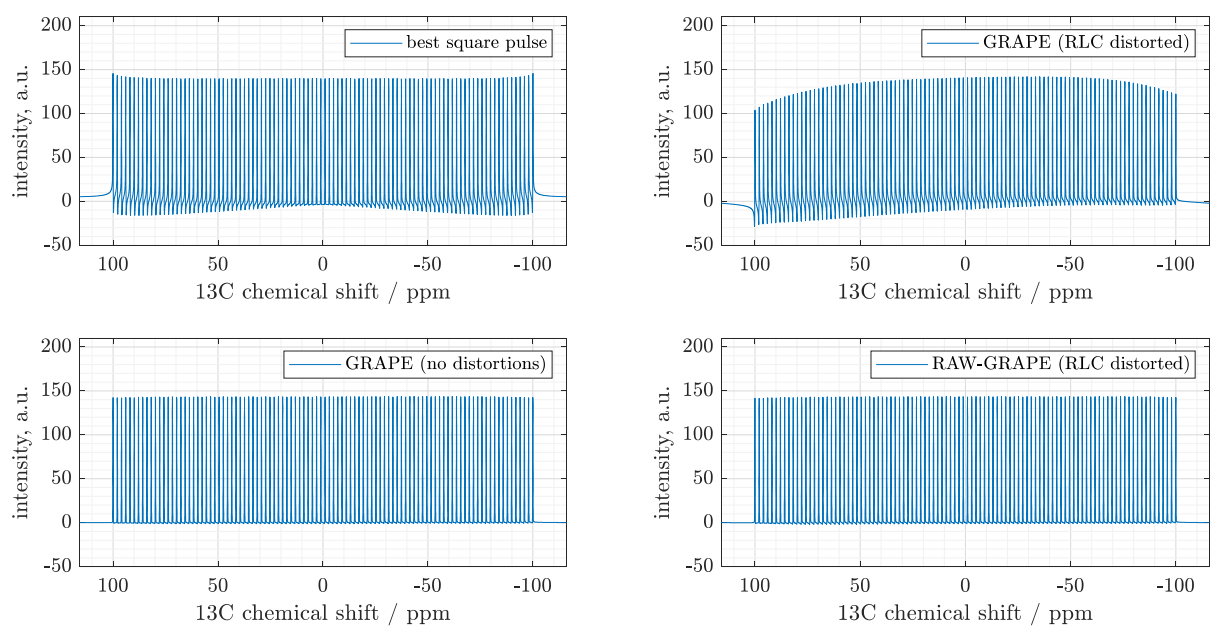

*Figure 3*. Performance comparison for the universal rotation pulses intended to accomplish the state space transformation in Eq (20) for an ensemble of 100 $^{13}$C nuclei spread uniformly over a ±100 ppm interval in a 28.18 Tesla (1.2 GHz proton frequency) NMR magnet. The shortest instrumentally available square pulse (**Top Left**) yields unavoidable phase distortions caused by the sinc bands of the pulse. An optimal control pulse designed using GRAPE is able to eliminate those distortions (**Bottom Left**), but the optimal pulse remains vulnerable to the effect of the RLC filter function of the probe circuit (Q=1000, **Top Right**). Using the RAW-GRAPE optimal control algorithm, where hardware distortion modelling is a part of the optimisation loop, eliminates the problem (**Bottom Right**). The script generating this figure is available in the example set supplied with versions 2.10 and later of Spinach.

An important aspect of quantum control is that different instruments and samples may have different distortion parameters. In bad cases, those parameters may not be the same from one experiment to the next, and may even vary across the sample – a famous case is $B_1$ field inhomogeneity within magnetic resonance coils and resonators[65]. It is therefore important to generate pulses that are resilient to ranges of instrumental distortion parameters. We have implemented this as an instance of ensemble control because ensemble handling is already available within *Spinach* library[51].

At the input syntax level, an ensemble of distortion cascades is supplied as multiple rows of the distortion function handle array, for example:

```
% Add ensemble of RLC distortions to the GRAPE workflow
control.distortion = { @(w)spf(w,p1(1)), @(w)spf(w,p2(1))
                       @(w)spf(w,p1(2)), @(w)spf(w,p2(2))
                       ...
                       @(w)spf(w,p1(n)), @(w)spf(w,p2(n)) };
```

where p1 and p2 are now arrays. This tells *Spinach* to create an ensemble of systems (parallelisation over ensembles is automatic) and to maximise the average fidelity across the ensemble.

Optimal control pulses generated using this procedure are resilient to ranges of distortion parameter values; this is illustrated in Figure 4 for two bi-parametric ensembles: power and RLC quality factor in the top panel, power and saturation level in the bottom panel.

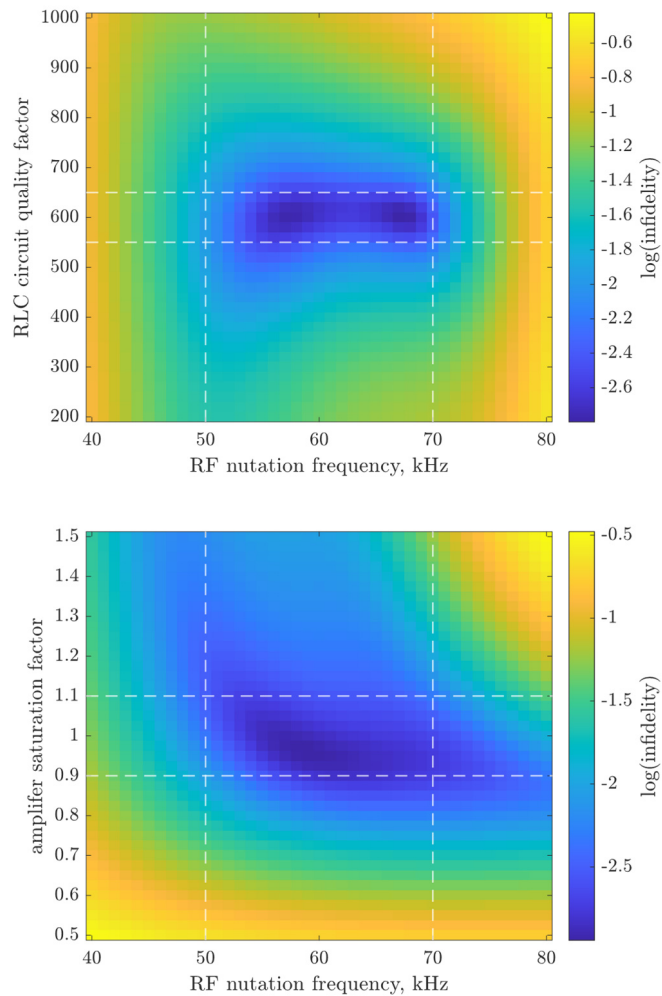

*Figure 4.* *Infidelity of the optimised universal rotation pulse (see Section 4 and Figure 3 caption for the parameters) in the presence of biparametric distortion ensembles. The pulse was optimised, using RAW-GRAPE, to be resilient to RF nutation frequency variations between 50 and 70 kHz, RLC quality factors variations between 560 and 640, and amplifier saturation levels between 50% to 150% of the maximum pulse power. Dashed lines in all panels indicate the distortion parameter ranges.* *(**Top Panel**) Biparametric ensemble of distortions (quality factor, nutation frequency) and the performance of the resulting pulse. (**Bottom Panel**). Biparametric ensemble of distortions (amplifier saturation level, nutation frequency) and the performance of the resulting pulse.*

Another important aspect of ensemble control is the trade-off between the fidelity that could have been achieved with fixed distortion parameters (sacrificed) and worst-case fidelity across the distortion parameter ensemble (improved). This is important for designing optimal control pulses that are transferable between instruments, samples, and measurement session. Figure 5 illustrates this principle: although RAW-GRAPE fidelity drops in the cases when distortions are benign, it does not then drop quite as much when distortions become significant.

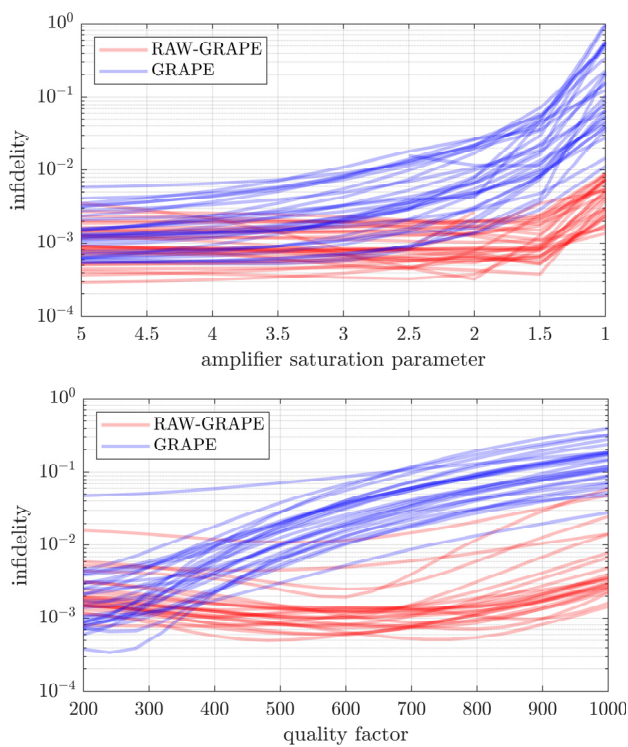

*Figure 5.* Performance illustrations (using "spaghetti plots" of multiple optimisation runs from different random initial guess pulses) for RAW-GRAPE (red lines) and original GRAPE (blue lines) in optimising a universal rotation pulse described in Section 4. (**Top Panel**) Ensemble of amplifier saturation levels, from negligible distortion when the ceiling is at 5 times the pulse power to strong distortion when the pulse is touching the ceiling. (Bottom Panel) Ensemble of NMR probe circuit quality factors, from negligible distortions at Q=200 (e.g. highly tuned room temperature probe) to strong distortion at Q=1000 (e.g. highly tuned specialised cryoprobe).

## 5. Conclusions and outlook

In the context of quantum optimal control, a significant problem is control sequence distortion by instrument electronics and optics[5,13,16,20,24,26,27]. Inverse transformations of those distortions may be ill-defined and unstable. However, inverting distortion functions is not actually necessary within the gradient ascent pulse engineering (GRAPE) framework for quantum optimal control – the only requirement is that a continuous Jacobian should exist; the rest of the mathematics is then reminiscent of the backpropagation algorithm used in artificial intelligence.

Using the algorithms described above, we have implemented arbitrary cascades of arbitrary differentiable distortions, and the corresponding ensemble control extensions, into version 2.10 of the open-source *Spinach* library, where the algorithm is called Response-AWare GRAPE (RAW-GRAPE for short). The user needs to supply function handles for the distortions; Jacobians are obtained internally using automatic differentiation. A set of common distortions (single-pole filter, single-zero filter, amplifier compression) is provided, those functions also serve as templates for customisation.

RAW-GRAPE improves robustness of optimal control sequences in situations where cascades and ensembles of instrumental distortions exist that are non-negligible and variable between instruments, samples, and measurement sessions.

## Acknowledgements

This work was supported by a research grant from the Centre for New Scientists at the Weizmann Institute of Science, by EPSRC (EP/Y035267/1) and MathWorks (who sponsored UR's PhD studentship), and used NVIDIA Tesla A100 GPUs through NVIDIA Academic Grants Programme. IK is indebted to the Weizmann Institute of Science for rescuing him from the crushing undergraduate teaching and administration workload at Southampton that would otherwise have made this work impossible.